\documentclass{aa}
\usepackage{graphicx}
\usepackage{natbib}
\usepackage[german,english]{babel}
\usepackage{lscape}

\newcommand{\gppr}{\stackrel{>}{\scriptstyle \sim}}
\newcommand{\gappr}{\raisebox{-0.4ex}{$\gppr$}}

\newcommand{\Porb}{\mbox{$P_\mathrm{orb}$}}

\newcommand{\Mwd}{\mbox{$M_\mathrm{wd}$}}
\newcommand{\Msec}{\mbox{$M_\mathrm{sec}$}}
\newcommand{\Rwd}{\mbox{$R_\mathrm{wd}$}}

\newcommand{\Twd}{\mbox{$T_\mathrm{wd}$}}

\newcommand{\Msun}{\mbox{$M_{\odot}$}}

\newcommand{\Mtr}{\mbox{$\dot{M}_{\mathrm{tr}}$}}
\newcommand{\Mcr}{\mbox{$\dot{M}_{\mathrm{cr}}$}}
\newcommand{\Macc}{\mbox{$\dot{M}_{\mathrm{acc}}$}}

\begin{document}

\title{Delays in dwarf novae {\rm\bf I}: The case of SS\,Cyg}
\authorrunning{Schreiber, Hameury, \and Lasota}
\author{Matthias R. Schreiber\inst{1}, Jean-Marie Hameury\inst{1}, \and Jean-Pierre Lasota\inst{2}}
\institute{ UMR 7550 du CNRS, Observatoire de Strasbourg, 11 rue
de l'Universit\'e, F-67000 Strasbourg, France \\
\email{mschrei@astro.u-strasbg.fr, hameury@astro.u-strasbg.fr}
\and
Institut d'Astrophysique de Paris, 98bis Boulevard Arago, 75014 Paris, France\\
\email{lasota@iap.fr} }
\offprints{M.R. Schreiber}

\date{Received 27 May 2003\,/\,Accepted 25 July 2003}

\abstract{
Using the disc instability model and a simple but
physically reasonable model for the X-ray, extreme UV, UV and
optical emission of dwarf novae we investigate the time lags
observed between the rise to outburst at different wavelengths. We
find that for ``normal'', i.e. fast-rise outbursts, there is good
agreement between the model and observations provided that the
disc is truncated at a few white dwarf radii in quiescence, and
that the viscosity parameter $\alpha$ is $\sim0.02$ in quiescence
and $\sim0.1$ in outburst. In particular, the increased X-ray flux
between the optical and EUV rise and at the end of an outburst,
is a natural outcome of the model. We cannot explain, however, the
EUV delay observed in anomalous outbursts because the disc
instability model in its standard $\alpha$-prescription form is
unable to produce such outbursts. We also find that the UV delay
is, contrary to common belief, slightly longer for inside-out
than for outside-in outbursts, and that it is not a good indicator
of the outburst type. 
}

\maketitle

\section{Introduction}

Dwarf novae are cataclysmic variables (CVs) in which the accretion
disc is subject to a thermal-viscous instability resulting in
recurrent outbursts lasting for a few days and separated by weeks
(see \citet{warner95-1} for an encyclopedic review of CVs and
\citet{lasota01-1} for a recent review of the disc instability
model). Their study has received considerable attention, since
these systems are often bright, usually vary on short
time scales and are relatively easy to observe so they are an
excellent testing ground for
the mechanisms transporting angular momentum in accretion
discs which, despite very significant recent progress, are not
yet fully understood \citep[see][for recent
reviews]{balbus+hawley98-1,balbus02-1}.

The disc instability model (DIM) is based on the existence of a
thermal-viscous instability in regions where hydrogen is partially
ionized, and opacities depend strongly on
temperature. If one plots a disc's thermal equilibria as the
effective disc temperature $T_{\rm eff}$ at a given radius $r$ (or
equivalently for viscous equilibria the mass transfer rate
$\dot{M}$) as a function of the disc surface density $\Sigma$, one
obtains the well known S-curve, in which the upper (hot) and
lower (cold) branches are stable and the intermediate one is
unstable. These branches are delimited by two critical values of
$\Sigma$, $\Sigma_{\rm max}$ above which no cool solution exists,
and $\Sigma_{\rm min}$ below which no hot solution is possible.

Provided that the mass transfer rate from the secondary
corresponds somewhere in the disc to this unstable branch, the
disc will not be steady. In quiescence, matter is accreted onto
the white dwarf at a rate lower than the rate at which it is
transferred from the secondary; the disc stays on the lower branch
of the S-curve. The disc mass therefore grows, and, at some point,
$\Sigma$ reaches $\Sigma_{\rm max}$; the temperature increases
locally, and a heating fronts forms that propagate towards the
white dwarf and towards the outer disc edge. Matter then
accretes faster than transferred from the white dwarf, the disc
empties, and a cooling front eventually starts from the outer edge
of the disc that brings the system into quiescence.

If the mass transfer rate is high, the accumulation time at the
outer disc edge can be shorter than the viscous diffusion time,
and the instability will be triggered in the disc outer regions;
the outburst is of the outside-in type. On the other hand, for low
mass transfer rates, the viscous time is the shortest, and the
outburst will be triggered at the inner edge; the outburst is of
the inside-out type. The limit between both types of outburst
therefore depends sensitively on parameters such as the mass
transfer rate, and the viscosity; it is therefore important to be
able to determine the type of observed outbursts for constraining
these parameters. Models predict that the outburst light curve is
asymmetric in the outside-in case, with a sharp rise and a longer
decay, whereas the rise and decay time should be of the same order
in the inside-out case. It is however not an easy task to
guess the outside-in nature of an outburst from its light curve:
models sometimes produce asymmetric outbursts which are of the
inside-out type \citep{buat-menardetal01-1}.

Another difference
that has often been put forward to distinguish both types of
outburst is the existence of a delay between the UV and the
optical rise during an outburst, the so-called UV delay which has
been measured for several
dwarf novae. As UV radiation is emitted
close to the white dwarf, one would naively expect that there
should be a long delay in cases where the outburst is triggered in
the disc outer regions, and no delay when the outburst starts at
the inner disc edge.

Recently, time lags similar to the UV delay have been observed
between the optical rise, an increase of X-ray emission and the
EUV light curve \citep{maucheetal01-1,wheatleyetal03-1}.
Especially the long orbital period, very bright
dwarf nova SS\,Cygni has intensively been monitored simultaneously at
different wavelengths. For four outbursts the EUV delay has been
measured and SS\,Cyg is the only dwarf nova system for which the
evolution of the X-ray emission throughout an outburst is
available. As in the case of the UV delay, differences in the
observed EUV delay have been interpreted as an indicator for the
outburst type \citep{maucheetal01-1,cannizzo01-1}.

There has been a long debate about the ability of the DIM to
reproduced observed UV delays. \citet{smak98-1} showed that the asserted
inability of the DIM to reproduce delays results from the
inadequacy of the numerical scheme used. The outburst type, and
hence the resulting delays, depend strongly on the modeling
itself; in particular, it is very difficult to obtain outbursts of
the outside-in type when the outer disc edge is kept fixed to a
given value \citep{hameuryetal98-1}. Similarly, the heating of the
outer regions by effects such as
the impact of the stream of matter flowing from the secondary, or
tidal torques dissipation also influences the nature of the
outbursts.

In this paper, contrary to the previous studies of the UV and
other delays, we use a version of the DIM
\citep{buat-menardetal01-1} which allows describing of the SS\,Cyg
outburst cycle with the real parameters (especially the large disc
size and mass-transfer rates corresponding to outside-in
outbursts) of this system. We use simple but reasonable
assumptions about the emission from the boundary layer to
investigate multi--wavelengths observations of this
observationally best studied dwarf nova. The paper is
organized as follows. Having reviewed the observations in Sect.\ref{s-obs}
we present our model in Sect.\,\ref{s-model} and its predictions for
SS\,Cygni in Sect.\,\ref{s-results}.
Finally
we compare our results with those of previous studies (Sect.\,\ref{s-prev})
and discuss them in the light of the observations (Sect.\,\ref{s-comp}).

\section{Reviewing the observations \label{s-obs}}

\begin{figure}
\includegraphics[width=8.5cm, angle=0]{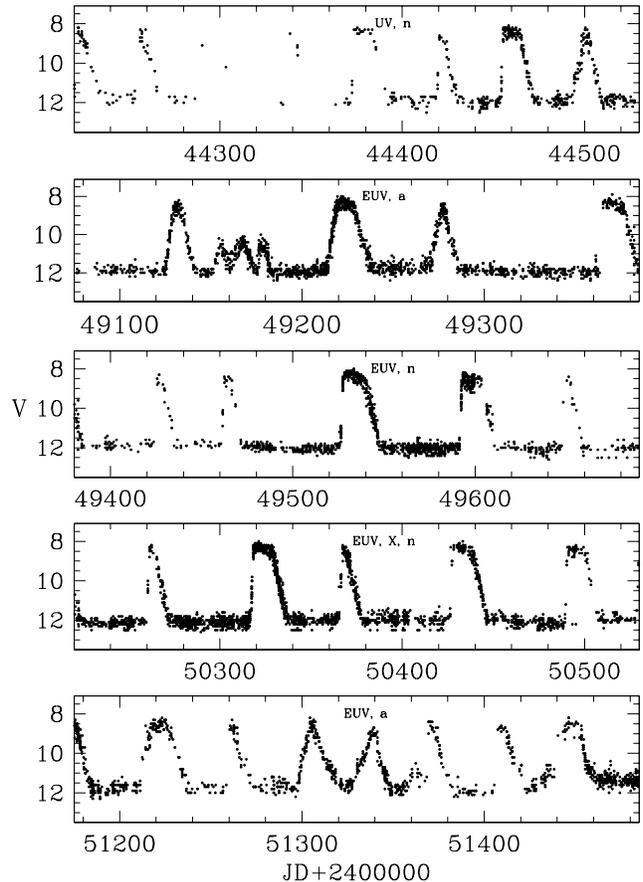}
\caption{\label{f-sscvis} Snapshots of the visual light curve of
SS\,Cyg. The five panels show parts of the light curve including
the outbursts for which observations in the UV, EUV, or X-ray
range exist (see Table\,\ref{t-obs1}). Outbursts called normal
(anomalous) are marked with n (a). The data is taken from the
AFOEV.}
\end{figure}

\begin{table*}
\newcounter{ref}
\newcommand{\tcite}{\stepcounter{ref}\arabic{ref}}
\newcommand{\tref}[1]{\stepcounter{ref}(\arabic{ref})\,\citealt{#1}}
\caption[]{\label{t-obs1} Outbursts of SS\,Cyg observed at
different wavelengths and the resulting delays. The quantities
$X_{\rm on}$ ($X_{\rm off}$) denote the time lag measured between
the optical rise and the increase (decrease) of the X-ray
emission.  The UV and EUV delays are measured at the beginning of
the optical rise (index 0) and at the half the maximum optical
flux (index 0.5) where also the width (W) of the outbursts is
measured. Outbursts with a slow optical rise are called
``anomalous'' (type a) whereas the optical flux increases rapidly
in the case of ``normal'' outbursts (type n).}
\setlength{\tabcolsep}{1.5ex}
\begin{tabular}{lccccccccccr}
\hline\noalign{\smallskip}
\hline\noalign{\smallskip}
\multicolumn{1}{c}{Instrument} &
JD+2440000 &
\multicolumn{1}{c}{$X_{\mathrm{on}}$}[d] &
\multicolumn{1}{c}{$X_{\mathrm{off}}$}[d] &
\multicolumn{1}{c}{$\Delta_{\mathrm{EUV,0}}$}[d] &
\multicolumn{1}{c}{$\Delta_{\mathrm{EUV,0.5}}$}[d] &
\multicolumn{1}{c}{$\Delta_{\mathrm{UV,0}}$}[d] &
\multicolumn{1}{c}{$\Delta_{\mathrm{UV,0.5}}$}[d] &
\multicolumn{1}{c}{W}[d] &
\multicolumn{1}{c}{Type} &
Ref.\\
\noalign{\smallskip}\hline\noalign{\smallskip}
Voyager      &  4372 &  -- & --  & --  & --         & $\sim$\,0.5 & $\sim$0.5 &
15 & n &\tcite\\
EUVE         &  9220 & --  & --  & 3.0 & $\sim$ 0.0 & --       & --   & 17 & a
& \tcite \\
EUVE         &  9530 & --  & --  & 1.5 & $\sim$ 1.5 & --       & --   & 15 & n
& 2 \\
EUVE \& RXTE & 10370 & 0.9 & 1.4 & 1.5 & $\sim$ 1.5 & --       & --   & 8 & n &2, \tcite, \tcite\\
EUVE         &  11335 & --  & -- & $>$ 5 & $\gappr$ 0.5 & --     & --   & 8 &
a & 2\\
\noalign{\smallskip}\hline\noalign{\smallskip}
\end{tabular}
\linebreak \setcounter{ref}{0} References:
\tref{cannizzoetal86-1}
\tref{maucheetal01-1},
\tref{wheatley00-1}, \tref{wheatleyetal03-1}
\end{table*}

SS\,Cygni has become the classical dwarf nova system for
analyzing the mechanism of dwarf nova outbursts
because a detailed visual long term light curve of rather regular
outburst behaviour is available (see Fig.\,\ref{f-sscvis} for
snapshots of the light curve and \citet[][]{cannizzo+mattei92-1}
for a detailed light curve analysis). In addition, SS\,Cyg has
been intensively observed at shorter wavelengths:

\noindent (1) In quiescence as well as during outburst
SS\,Cyg is -- as dwarf novae in general -- a source of X-ray
emission. Spectral fits to the X-ray flux observed in quiescence
suggest a Bremsstrahlung origin from gas with temperatures of
$2-20$\,keV and luminosities of the order of a few
$10^{32}$\,d$_{100}^2$\,erg\,s$^{-1}$ where d$_{100}$ is the
distance in units of 100\,pc
\citep{yoshidaetal92-1,done+osborne97-1,ponmanetal95-1}. Relating
this X-ray luminosity to the emission from the boundary layer
requires an accretion rate of
$\Macc\sim\,10^{15}$\,d$_{100}^2$\,g\,s$^{-1}$, assuming
gravitational energy conversion onto a 1.0 \Msun white dwarf.

\noindent
(2) \citet{cannizzoetal86-1} present simultaneous observations in the UV and
optical during an outburst and find the rise of the UV flux being delayed to
the optical by $\sim0.5$\,days ($\Delta_{\mathrm{UV,0}}$ and
$\Delta_{\mathrm{UV,0.5}}$ in Table\,\ref{t-obs1}).

\noindent (3) Further simultaneous observations of SS\,Cyg at
optical and EUV wavelengths have been published for four outbursts
of SS\,Cygni \citep{maucheetal01-1}. Two of these outbursts are
called ``normal'' outbursts (marked with n in Fig\,\ref{f-sscvis}
and Table\,\ref{t-obs1}) which refers to the observed fast optical
rise.  The time lag between the optical and the EUV is
$\Delta_{\mathrm{EUV,0}}\approx \Delta_{\mathrm{EUV,0.5}}
\sim\,1.5$\, days for these outbursts, i.e. the EUV delay is 
almost independent on where
one measures the delay. In contrast, the EUV delay measured for the
two ``anomalous'' outbursts (i.e. outbursts with a slow optical
rise, denoted with a in Fig.\,\ref{f-sscvis} and
Table\,\ref{t-obs1}) is essentially longer at the onset of the
optical rise but decreasing with increasing visual brightness 
(see $\Delta_{\mathrm{EUV,0}}$ and
$\Delta_{\mathrm{EUV,0.5}}$ in Table\,\ref{t-obs1}).

\noindent (4) \citet{wheatley00-1} and
\citet{wheatleyetal03-1} present simultaneous observations at
optical, X-ray, and EUV wavelengths and finds: the hard X-ray flux
increases 0.9\,d ($X_{\rm on}$ in Table \ref{t-obs1}) after the optical and
half a day
before the EUV rise, at which point X-rays are abruptly shut off. $X_{\rm
off}$ in Table \ref{t-obs1} denotes the time lag between the optical rise and
the sudden decrease of X-ray emission.
At the end of the optical decline the X-ray emission rises again
for approximately two days.

Snapshots of the optical light curve corresponding to the five
outbursts for which simultaneous multi--wavelength observations
exist (Table\,\ref{t-obs1}) are shown in Fig.\,\ref{f-sscvis}.

\section{The model \label{s-model}}

We use here the version of the DIM described in
\citet{hameuryetal98-1} in which heating of the disc by the tidal
torque and the stream impact have been incorporated
\citep{buat-menardetal01-1}. We detail below the contribution
of each constituent of the system to the light curve at various
wavelengths; as we shall see, the optical light curve is dominated
by the accretion disc, with some contribution from the hot spot
and the secondary star; the UV in outburst originates from the
disc, and the boundary layer is the source of hard radiation.

\subsection{Contribution of the boundary layer}

The transition region between the accretion disc and the white
dwarf, the boundary layer is generally thought to be a source of UV, EUV
and hard X-ray photons.
Approximately one half of the accretion energy should be released in this
region.

Although the detailed nature of this region is very uncertain,
simple models
\citep{pringle+savonje79-1,tylenda81-1,patterson+raymond85-1} as
well as more detailed calculations \citep{narayan+popham93-1}
suggest the transition from an optically thin X-ray emitting
region to an optically thick boundary layer when the rate at which
mass is supplied to the boundary layer from the accretion disc
exceeds a certain value around $\,10^{16}$g\,s$^{-1}$. Hence, for
high accretion rates the optically thick boundary layer is
expected to dominate the EUV and soft X-ray emission of dwarf
novae whereas for low accretion rates the emission of X-rays is
expected.

This picture is not only favoured by pure theoretical arguments but it appears
also reasonable considering the observations reviewed in the previous
section.
We therefore assume here
that the boundary layer is optically thin if the mass accretion rate is
below $\dot{M}_{\mathrm{cr}}=10^{16}$gs$^{-1}$
and optically thick otherwise.
We assume that the optically thin boundary
layer emits hard X-rays proportional to the accretion rate
below $\dot{M}_{\mathrm{cr}}$:
\begin{equation}
L_{\mathrm{X}}=L_{\mathrm{BL}}=\frac{G\dot{M}_{\mathrm{acc}}\Mwd}{2\Rwd}.
\end{equation}
The optically thick boundary layer is approximated with a blackbody of
the effective temperature
\begin{equation}
T_{\mathrm{BL}}^4=\frac{L_{\mathrm{BL}}}{f_{\mathrm{em}}\sigma\,4\pi\,\Rwd^2}=\frac{G\dot{M}_{\mathrm{acc}}\Mwd}{f_{\mathrm{em}}\sigma\,4\pi\,\Rwd^2\,2\Rwd},
\label{eq:twd}
\end{equation}
where $f_{\mathrm{em}}$ is a parameter representing the fractional emitting
region of the boundary layer. Clearly, the boundary layer will expand with
increasing accretion rate,
i.e. $f_{\mathrm{em}}$ increases with
$\dot{M}_{\mathrm{acc}}$.
We approximate the expansion of the boundary layer with
$\dot{M}_{\mathrm{acc}}$
following \citet{patterson+raymond85-2}:
\begin{equation}
f_{\mathrm{em}}=10^{-3}\left(\frac{\dot{M}_{\mathrm{acc}}}{10^{16}\mathrm{gs}^{-1}}\right)^{0.28}.
\label{eq:deff}
\end{equation}
with $\dot{M}_{\mathrm{acc,16}}=\dot{M}_{\mathrm{acc}}/10^{16}$gs$^{-1}$.

We are aware of the approximative nature of the above prescription as
the choice of $\dot{M}_{\mathrm{cr}}$ is somewhat arbitrary and we do not take
into account the time it may take the boundary layer to switch between the 
optically thin
and optically thick state.
However, observations (see point 4 in the previous section) indicate that the
transition is rather immediate and for the purpose of this paper our simple
approach is sufficient.

\subsection{Thermal emission from the white dwarf}

The white dwarf contributes to the total emission,
both in the UV and optical. Throughout this paper we assume that
the effective temperature of the white dwarf is $\Twd=18\,000$\,K.
We neglect white dwarf cooling and
assume that $\Twd$ remains constant; this is justified as long as
one is interested only in the initial phases of the outburst (in
systems in which white dwarf cooling has been observed, the
cooling time is larger than 10 days, \citet{sion99-1}). In view of
the uncertainties associated with the treatment of the
boundary layer,
it is sufficient to assume that the spectrum is
that of a blackbody.

\subsection{The secondary star}

In the following (except when otherwise stated) we consider that
the secondary is a 0.7 \Msun main sequence star, with effective
temperature 4000\,K. The spectrum is taken from \citet{kurucz93-1}.
During an outburst, the effective temperature of the
irradiated hemisphere of the secondary increases, and the spectrum
of the secondary becomes the sum of two blackbodies with different
effective temperatures, but the same emitting area, provided the
luminosity is averaged over the orbital period. We assume here
that:
\begin{equation}
T_2^4 = T_*^4 + \left( {\Rwd \over a} \right)^2\left\{T_{\rm BL}^4 f_{\rm em} +\Twd^4 \right\}
\label{eq:ts}
\end{equation}
where $T_2$ and $T_*$ are the effective temperatures of the
illuminated and un illuminated hemispheres respectively, $T_{\rm
BL}$ is given by Eq. (\ref{eq:twd}), and $a$ is the orbital
separation. Note that the disc luminosity does not enter in Eq.
(\ref{eq:ts}), because its luminosity is emitted perpendicular to
the orbital plane, and only a small fraction of it can effectively
heat the secondary.

\subsection{The hot spot}
In our model, a fraction of the energy released by the impact of
the stream onto the disc is assumed to be thermalized, and is
already included in the disc model. The remaining fraction is
released in the hot spot, for which we assume a blackbody spectrum
with effective temperature 10,000 K. In the following we assume
that one half of the stream impact energy is emitted by the hot
spot.

\subsection{Disc emission}

The local spectrum of the disc is assumed to be given by
\citet{kurucz79-1,kurucz93-1}. This assumption is probably not very
good, especially during quiescence, when the optical depth of the
disc is not large. Both the spectra observed during quiescence in
e.g. HT Cas \citep{vrielmannetal02-1} and the predicted ones
\citep{idanetal99-1} differ from the simple stellar
spectra. For comparison, and also to get some hint on the
uncertainty linked to the precise modeling of the spectrum, we
also calculate the disc spectrum by summing blackbodies.

\begin{table}
\begin{center}
\caption[]{\label{t-ssc_par} Binary parameter of SS\,Cyg.}
%
\setlength{\tabcolsep}{6ex}
\begin{tabular}{lc}
\hline\noalign{\smallskip}
\hline\noalign{\smallskip}

\Porb/hr       & 6.6\\
$\Mwd/\Msun$ & 1.19 \\
$\Msec/\Msun$ & 0.70\\
$<R_{\mathrm{out}}>/10^{10}$cm & 5.4\\
\Rwd/$10^8$\,cm & 3.9 \\

\noalign{\smallskip}\hline\noalign{\smallskip}
\end{tabular}
\linebreak
\end{center}
\end{table}

\section{Results \label{s-results}}
In this section we present the results we obtain applying the
model to a system with the orbital parameter of SS\,Cyg given in
Table\,\ref{t-ssc_par}. The mean outer radius $<R_{\mathrm{out}}>$
is taken to be the average of $r_1$, $r_2$, and $r_{\mathrm{max}}$
calculated in Table\,1 of \citet{paczynsky77-1}.

We calculated monochromatic light curves at three wavelengths
($\lambda=(100,\,1250,\,5500)\,$\AA)
representative for the EUV, UV, and optical flux. For low
accretion rates we additionally assumed the energy released in the
boundary layer being related to X-ray emission. In
Table\,\ref{t-sample} we have collected the results of our
comprehensive numerical investigation. In particular we studied
the influence of three so far rather unconstrained ingredients of
the model, i.e. the inner boundary condition, the viscosity
parameter $\alpha$, and the mass transfer rate from the secondary.

In general, the obtained UV emission is rising after the optical
flux but before the EUV increases. We find that the delays
depend strongly on where we measure them. In Table\,\ref{t-sample}
we therefore give values for time lags at the onset of the
optical rise (index 0) and at half the maximum optical emission 
(index 0.5).

\begin{center}
\begin{table*}
\caption[]{\label{t-sample}Parameter and delays of calculated
outbursts. The obtained delays depend on where one measures them;
for the UV and EUV we give values at the beginning of the optical rise
(index 0) and at half the maximum optical emission (index 0.5) where also 
the width
of the outbursts (W) has been measured. $R_{\rm init}$ denotes the radius
where the heating front started. $X_{\rm on}$ is the time lag between the
optical rise and increased X-ray emission. The time $X_{\rm off}$ after the
optical rise the boundary layer becomes optically thick and the expected
X-rays are shut off.}
\setlength{\tabcolsep}{2.2ex}
\begin{tabular}{lllllllllllllr}
\hline\noalign{\smallskip} 
\hline\noalign{\smallskip} 
model &
\multicolumn{1}{c}{$\dot{M}_{\mathrm{tr}}$}&
\multicolumn{1}{c}{$\alpha_{\mathrm{h}}$}&
\multicolumn{1}{c}{$\alpha_{\mathrm{c}}$}&
\multicolumn{1}{c}{$\mu_{30}$}&
\multicolumn{1}{c}{$X_{\mathrm{on}}$} &
\multicolumn{1}{c}{$X_{\mathrm{off}}$} &
\multicolumn{1}{c}{$\Delta_{\mathrm{EUV,0}}$} &
\multicolumn{1}{c}{$\Delta_{\mathrm{EUV,0.5}}$} &
\multicolumn{1}{c}{$\Delta_{\mathrm{UV,0}}$} &
\multicolumn{1}{c}{$\Delta_{\mathrm{UV,0.5}}$}&
\multicolumn{1}{c}{$R_{\mathrm{init}}$} &
\multicolumn{1}{c}{W} \\
& \multicolumn{1}{c}{[$10^{16}{\rm gs}^{-1}$]}& & & &
\multicolumn{1}{c}{[d]} & \multicolumn{1}{c}{[d]} &
\multicolumn{1}{c}{[d]} & \multicolumn{1}{c}{[d]}&
\multicolumn{1}{c}{[d]} & \multicolumn{1}{c}{[d]}&
\multicolumn{1}{c}{[$10^{10}$\,cm]}&
\multicolumn{1}{c}{[d]}\\
\noalign{\smallskip}\hline\noalign{\smallskip}
1 & 8.5       & 0.1 & 0.02 & 0 & 0.3 & 0.9 & 1.0 & 1.4 & 0.3 & 0.8 & 0.85 & 7    \\
2 & 10        & 0.1 & 0.02 & 0 & 0   & 1.5 & 1.6 & 1.4 & 1.0 & 0.7 & 0.06 & 14   \\
3 & 12.5      & 0.1 & 0.02 & 0 & 0   & 2.4 & 2.6 & 1.5 & 1.8 & 1.0 & 0.06 & 23   \\
4 & 13.75     & 0.1 & 0.02 & 0 & 0.4 & 1.0 & 1.1 & 1.5 & 0.4 & 0.9 & 1.26 & 25   \\
5 & 15        & 0.1 & 0.02 & 0 & 0.45 & 1.0 & 1.2 & 1.5 & 0.4 & 0.9 & 1.35  & 36 \\
\noalign{\smallskip}\hline\noalign{\smallskip}
6 & 11        & 0.1 & 0.02 & 2 & 0   & 0.8 & 0.8 & 1.6 & 0.7 & 1.0 & 0.25 & 16 \\
7 & 13.75     & 0.1 & 0.02 & 2 & 0   & 0.8 & 1.2 & 1.5 & 0.8 & 0.9 & 0.25 & 24 \\
8 & 15        & 0.1 & 0.02 & 2 & 0.3 & 0.7 & 0.9 & 1.4 & 0.3 & 0.7 & 1.2 & 37  \\
\noalign{\smallskip}\hline\noalign{\smallskip}
9 & 15        & 0.2 & 0.02 & 2 & 0.2 & 0.4 & 0.5 & 0.6 & 0.1 & 0.2 & 1.40 & 20  \\
10 & 15        & 0.1 & 0.01 & 2 & 0.3 & 0.7 & 0.8 & 1.0 & 0.1 & 0.3 & 1.31 & 29  \\
\noalign{\smallskip}\hline\noalign{\smallskip}
11 & 14.1$^{*}$& 0.1 & 0.02 & 2 & 0   & 0.8 & 0.8 & 1.4 & 0.8 & 1.0 & 0.25 & 9 \\
& 14.1$^{*}$& 0.1 & 0.02 & 2 & 0   & 0.8 & 0.8 & 1.4 & 0.8 & 0.9 & 0.25 & 60 \\
& 14.1$^{*}$& 0.1 & 0.02 & 2 & 0.3 & 0.7 & 0.8 & 1.3 & 0.3 & 0.7 & 1.28 & 6 \\
& 14.1$^{*}$& 0.1 & 0.02 & 2 & 0.3 & 0.7 & 0.8 & 1.2 & 0.3 & 0.7 & 1.32 & 51  \\
\noalign{\smallskip}\hline\noalign{\smallskip}
\end{tabular}\linebreak
*: variable mass transfer rate ($\pm$ 15\%, see text)\\
\end{table*}
\end{center}

\begin{figure}
\begin{center}
\includegraphics[width=8.5cm, angle=0]{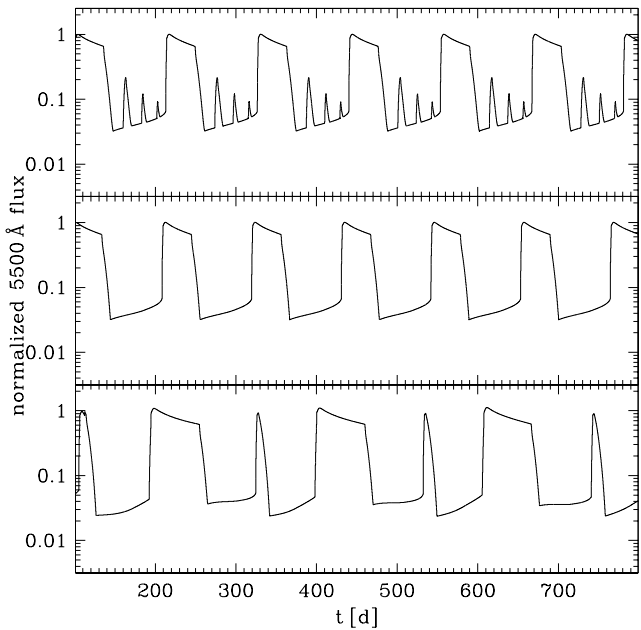}
\caption{\label{f-cal6}Calculated long term light curves. From
top to bottom: without truncation (model 5), with truncation
(model 8), and finally with truncation and assuming that the mass
transfer rate varies slightly and smoothly (model 11).}
\end{center}
\end{figure}

\subsection{Fixed inner boundary}

Assuming that the disc extends down to the surface of the white
dwarf, i.e. $R_{\mathrm{in}}=3.9\times\,10^{8}$ we performed
calculations assuming five different mass transfer rates (model
1--5 in Table\,\ref{t-sample}). As in earlier calculations
\citep[e.g.][]{hameuryetal98-1} we find in this case the
calculated light curve cycles consisting of several short low
amplitude outbursts followed by a larger one (see Fig.
\ref{f-cal6}). This form of the DIM cannot reproduce the
observed light-curve of SS\,Cyg and similar dwarf novae but it can
be used to study various density and temperature distributions in
the pre-outburst disc. The delays listed in Table\,\ref{t-sample}
refer to the long outbursts.

\begin{figure*}
\begin{center}
\includegraphics[width=10cm, angle=270]{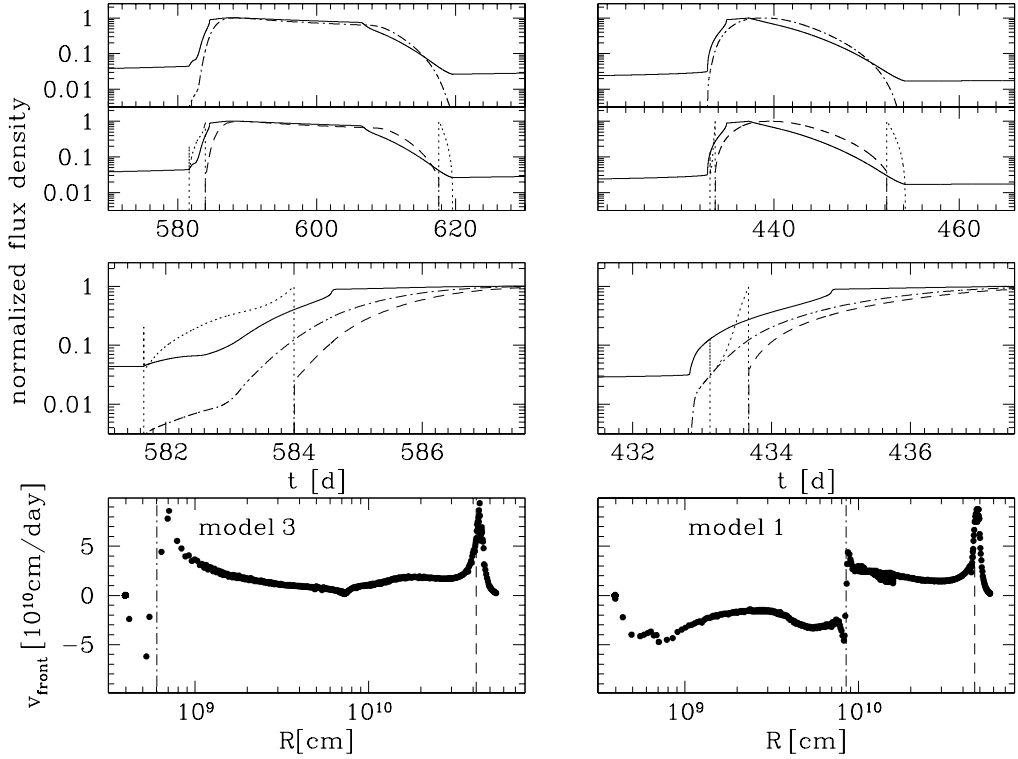}
\caption{\label{f-cal1} The predicted normalized optical (solid
line) and UV flux (dashed--dotted) as well as the normalized
boundary layer emission, i.e. X-rays (dotted) and EUV (long
dashed) for inside-out (left) and outside-in (right)
outbursts. The intermediate panel shows a detailed view of the
outburst rise, and the scale is the same for both models. The
bottom panels show the heating front velocity corresponding to the
rise of the outbursts. The ignition radius is marked by the
vertical dashed--dotted line. The long dashed line on the right
gives the position of the disc radius during quiescence. Having
arrived at the outer edge, the plotted velocity gives no longer
the speed of the heating front but that of the expansion of the
disc.}
\end{center}
\end{figure*}

As expected, the width of these outbursts increases with the mass
transfer rate. For high mass transfer rates the accumulation time
scale becomes shorter than the viscous diffusion time scale
leading to outbursts of the outside--in type; conversely,
inside-out outbursts are expected at low mass transfer rates. An
exception from this rule is model 1, i.e. the calculation with the
lowest mass transfer rate. This happens because the wide
outburst is preceded by a large number of low-amplitude inside-out
outbursts during which the heating front dies out before reaching
the outer regions allowing mass to accumulate there.
Fig.\,\ref{f-cal1} shows the normalized flux densities at
the relevant wavelengths and the heating front velocities for two
calculations (model 1,3) with the disc extending down to the white
dwarf.

\subsubsection{Inside-out outbursts}

At first we consider the properties of {\em inside--out} outbursts
(model 2,3; left panels of Fig.\,\ref{f-cal1}): The optical rise
begins immediately when the heating front starts even though
this happens close to the white dwarf. With the heating front
reaching the outer regions of the disc, the optical rise gets
faster and reaches its maximum value when the outer edge of
the disc becomes hot and the disc is expanding radially. At large
radii the velocity of the heating front increases due
to the effects of additional heating of the outer edge
(Fig.\,\ref{f-cal1}). The UV emission increases immediately
after the ignition of the inside--out heating front but the
accretion rate in the inner regions of the disc remains rather
low. Hence, UV does not even reach one percent of its maximum
value during the early stages of the optical rise. The situation
is similar but even more drastic for the EUV emission. The mass
accretion rate onto the white dwarf stays far below $\Mcr$ until
the heating front has reached the outer disc regions.
This explains the long delays at the onset of the optical rise
i.e. $\Delta_{\mathrm{UV,0}}$ and $\Delta_{\mathrm{EUV,0}}$. When
the heating front has reached more massive outer disc
regions, the UV and the EUV emission are rising faster,
leading to somewhat decreased delays $\Delta_{\mathrm{UV,0.5}}$
and $\Delta_{\mathrm{EUV,0.5}}$. Neither the UV nor the EUV reach
their maximum value before the disc has adjusted to the quasi
stationary outburst state. Concerning the X-ray emission our
calculations predict that the emission from the optically thin
boundary layer rises immediately when the inside--out heating
front is triggered. The slow increase of the mass accretion rate
is displayed by the relatively long lasting X-ray emission during
the early rise of the outburst (dotted line in the left panel of
Fig.\,\ref{f-cal1}).

\subsubsection{Outside-in outbursts}

The overall situation is very different for {\em outside--in}
outbursts where the heating front ignites far away from the white
dwarf (model 1, 4 and 5). One should stress here that while
inside-out outbursts always start close to the inner disc edge,
the outside-in outbursts may start relatively far from the outer
rim; as a result two heating fronts propagate and such
outbursts are not really purely ``outside-in'' \citep[see]
[for examples and discussion]{buat-menardetal01-1}.

When the
front reaches much earlier the outer parts of the disc the optical
flux increases sharply immediately after the instability has
been triggered. In addition the entire optical rise is shorter as
there are two heating fronts in the disc; the outside--in
heating front is indeed the fastest one, but both fronts are
faster than a pure inside--out heating front (see the bottom
panels of Fig.\,\ref{f-cal1}). Naively one could expect the UV and
EUV delays to be longer than for inside-out outbursts because
of this fast optical rise and because the heating front has to
reach the inner disc before the UV and EUV fluxes significantly
rise. However, the opposite is true. In addition to
differences in their propagation direction, outside--in and
inside--out fronts differ in many other respects.
Inside-out heating fronts must  propagate ``uphill'' against the
surface-density and angular-momentum gradients. This makes the
propagation difficult and such fronts can be subject to dying before
reaching the outer disc
\citep[see][for a detailed
discussion]{linetal85-1,menouetal99-2,lasota01-1}. In contrast, an outside--in
heating front starts in high surface density regions and has an easy way
``downhill'' sliding down along the gradients. They never die
before fulfilling their task.
Thus, for an outside--in front the accretion rate in the inner
regions of the disc as well as the accretion rate onto the white
dwarf increase much more rapidly than in the case of an
inside--out heating front. As a consequence, the UV and EUV delay
at the onset of the optical rise are significantly {\em shorter}
than in the inside--out case. As the inward moving transition
front is faster, the final stages of the rise are dominated by the
inside--out part of the heating front and by the
adjustment of the disc to the quasi stationary state. Hence, the
delays measured closer to the maximum ($\Delta_{\mathrm{UV,0.5}}$,
$\Delta_{\mathrm{EUV,0.5}}$) are nearly independent on the
ignition radius.

The early rise of the accretion rate which is related to
the X-ray flux is delayed relative to the optical by $0.3-0.45$\,days, i.e.
the time it takes the heating front to reach the inner edge (see
Fig.\,\ref{f-cal1},\,\ref{f-cal2}).

\subsubsection{Decline and quiescence}

Clearly, during the late optical decline of every outburst our model
predicts an increase
of the X-ray emission when the mass accretion rate decreases below
$10^{16}$\,g\,s$^{-1}$ and the boundary layer is expected to become optically
thin.
Cooling fronts are slower than outside--in heating fronts but
their velocity is comparable to that of inside--out heating fronts at small
radii \citep[see][for a detailed study of the properties of transition
fronts]{menouetal99-2}.
Consequently, the predicted duration of the
increased X-ray flux (in the following called the ``X-ray on''
state) is similar to that at the rise of an
inside--out outburst (i.e. $\sim\,2$\,days).

Finally we want to place emphasis on the predicted accretion rate
during quiescence which is in the range of
$10^{11}-10^{13}$\,g\,s$^{-1}$ for both outburst types.
Notice, although the duration of the ``X-ray on'' state corresponds to
the observed one, the calculated accretion rates during quiescence are
several orders of magnitude lower than those deduced from observations.
This will be discussed in more detail in Sect.\,\ref{s-comp}.

\subsection{Truncation of the inner disc}

\begin{figure*}
\begin{center}
\includegraphics[width=10cm, angle=270]{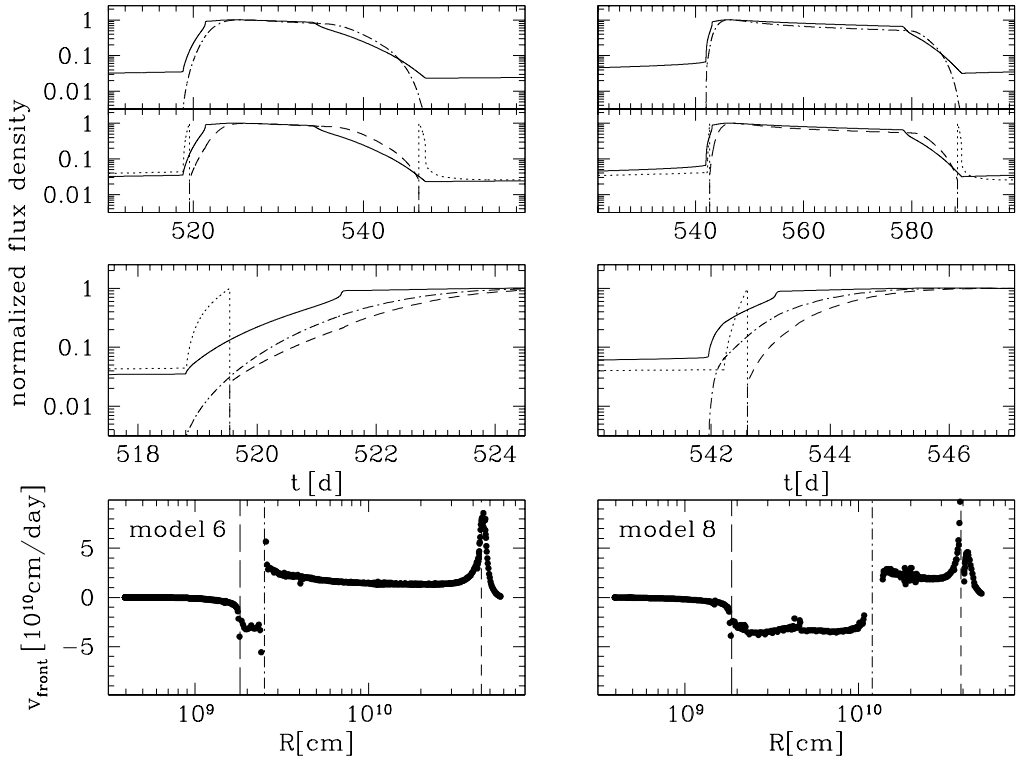}
\caption{\label{f-cal2}Same as Fig.\,\ref{f-cal1} but assuming the
inner disc being truncated. Around $R=2\times10^{9}$\,cm the
heating front reaches the inner edge of the disc indicated
by the long dashed vertical line.
Then the velocity
decreases from the speed of the heating front to that of the
inward expansion of the disc governed by Eq.\,(\ref{eq-rin}).
Note that the scale of the intermediate panel is the same as
in Fig. \ref{f-cal1}.}
\end{center}
\end{figure*}

During the last decade it has been intensively discussed whether
an optically thick accretion disc
extends down to the white dwarf
in quiescent dwarf novae. The main reason for this
discussion was the serious discrepancy between calculated accretion rates and
those deduced from X-ray observations. Disc truncation
by evaporation \citep{meyer+meyer-hofmeister94-1} and/or by a
magnetic field \citep{lasotaetal95-1,hameuryetal97-1} could bring
into agreement models and observations.
Also the alleged difficulties of the DIM with the UV-delay
motivated interest in truncated discs.
\citet{king97-1} suggested that irradiation of the disc
by the white dwarf can keep the inner disc in the hot ionized state
leading to very low surface densities
in this region\footnote{Disc irradiation does not
however have the same
consequences as plain disc truncation; for a detailed analysis of
disc irradiation see
\citet{hameuryetal99-1,stehle+king99-1,schreiber+gaensicke01-1}.}
and \citet{livio+pringle92-1} analyzed the effects of a weakly magnetized 
white dwarf.   

To discuss in detail each mechanism which has been proposed in the context of 
SS\,Cygni is beyond the scope of this paper. 
For our purpose it is sufficient
just to assume the formation of an inner hole during quiescence.
We assume here that the inner disc radius is given by the
``magnetospheric'' radius:
\begin{equation}
R_{\mathrm{in}}= R_{\mathrm{M}}= 9.8\times10^8
\dot{M}_{15}^{-2/7}\Mwd^{-1/7}\mu_{30}^{4/7}\mathrm{cm}
\label{eq-rin}
\end{equation}
where $\mu_{30}$ is the magnetic moment of the white dwarf in
units of $10^{30}$\,G\,cm$^3$
\citep[][]{hameury+lasota02-1}.

Clearly, because the accretion flow is disrupted by the magnetic
field, a boundary layer will not form when the disc is truncated.
Instead, matter flows along the field lines, and its kinetic
energy is released in a shock at the surface of the white dwarf.
The spectral range of this emission depends on whether the flow is
optically thin or not; for the sake of simplicity, we use the same
condition as for the boundary layer case. This is an apparently
crude approximation. Assuming, however, as an antithesis that the
flow becomes optically thick precisely when the disc reaches the
surface of the white dwarf would be equivalent to a small increase
of \Mcr\ because the accretion rate onto the white dwarf exceeds
$10^{16}$g\,s$^{-1}$ shortly before the disc reaches the white
dwarf. Such a small increase of \Mcr\ would lead to somewhat
higher values for the predicted duration of the ``X-ray on'' state
and the early EUV delay ($\Delta_{\rm EUV,0}$) but not affect the
conclusions of this paper.

Table\,\ref{t-sample} lists
the delays we obtained for three different mass transfer rates and
$\mu_{30}=2$. Fig.\ref{f-cal2} shows the normalized flux densities and
heating front velocities for two outbursts. The optical rise for
inside--out outburst is faster than without truncation as the
heating front is forced to start at $R_{\mathrm{in}}>\Rwd$, i.e.
in a region with higher surface density (Fig.\,\ref{f-cal2}).
Therefore, the delays $\Delta_{\mathrm{UV,0}}$ and
$\Delta_{\mathrm{EUV,0}}$ are shorter than for inside--out
outbursts without truncation. The predicted EUV delay of
outside--in and inside--out outbursts becomes comparable whereas
$\Delta_{\mathrm{UV,0}}$ remains significantly shorter for
outside--in outbursts.

\begin{figure*}
\begin{center}
\includegraphics[width=10cm, angle=270]{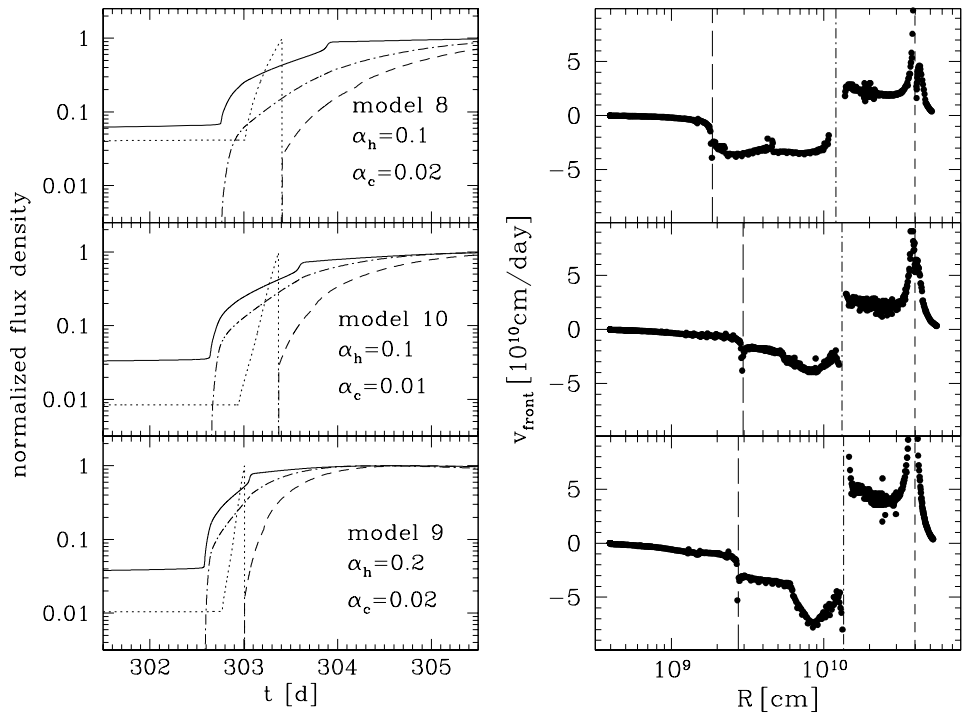}
\caption{\label{f-cal3} The same as Figs.\,\ref{f-cal1} and \ref{f-cal2} but
for different values of $\alpha$. From top to bottom the plots correspond to
model 8, 10, and 9 (see Table\,\ref{t-sample} for the predicted delays).}
\end{center}
\end{figure*}

Truncation is also changing the situation for the early rise of
the accretion rate which is correlated to the predicted hard X-ray
flux. Considering the arrival of the heating front at the inner
edge two effects have to be mentioned. First, the time it
takes outside--in heating fronts to reach the inner edge decreases
slightly because the heating front just has to reach
$R_{\mathrm{in}}= R_{\mathrm{M}}>\Rwd$. Second, once the
heating front reached $R_{\mathrm{in}}$ the disc starts filling the inner hole,
i.e. $R_{\mathrm{in}}$ is decreasing.
This process is rather slow compared to the heating front
velocity (see Fig.\,\ref{f-cal2}). The increase of the mass
accretion rate is therefore less sudden and the small spike seen
in Fig.\,\ref{f-cal1} disappears. The expected X-ray emission
at the end of an outburst is also influenced by the presence of
truncation. The initial velocity of cooling fronts is rather high
but as it propagates inward it relaxes to an essentially lower
speed which depends just on the location of the front
\citep{menouetal99-2}. Consequently, if the inner disc is
truncated, the cooling front is reaching earlier the inner edge
which reduces the duration of the ``X-ray on'' state at the end of
an outburst.

A final note on the effects of truncation concerns the accretion
rate during quiescence:  the postulation of an inner hole
dramatically increases the predicted X-ray flux during quiescence
(compare Fig.\,\ref{f-cal1} and \ref{f-cal2}).
Instead of $10^{11}-10^{13}$\,g\,s$^{-1}$ without truncation we obtain now
$\Macc\sim\,3-5\times10^{14}$\,g\,s$^{-1}$.
This increase of the expected X-ray emission during quiescence results from
the fact that the DIM predicts accretion rates which increase with radius
while the disc accumulates mass.

We want to stress that assuming evaporation (e.g. together with a radially
extended optically thin boundary layer
\citep{narayan+popham93-1,medvedev+menou02-1}) instead of a weakly magnetic
white dwarf would lead to almost identical results.

\subsection{Changing $\alpha$ \label{s-alpha}}

The third part of Table\,\ref{t-sample} lists results obtained with different
values of $\alpha$.
As long as this parameter is physically rather unconstrained, analyzing the DIM
requires discussing different values of $\alpha$.
We calculated again
light curves with $\mu_{30}=2$
but increased (decreased) $\alpha_{\mathrm{h}}$ ($\alpha_{\mathrm{c}}$).
As we choose a high mass transfer rate ($\Mtr=15\times10^{16}$\,g\,s$^{-1}$)
the outbursts are of the outside--in type.

Increasing $\alpha_{\mathrm{h}}$ (model 9, Fig.\,\ref{f-cal3} bottom) or
decreasing $\alpha_{\mathrm{c}}$ (model 10, Fig.\,\ref{f-cal3} middle) leads
to smaller accretion rates during
quiescence.
This is because the accretion rate on the lower branch of the S-curve
corresponding to $\Sigma_{\rm min}$
decreases when the ratio $\alpha_{\mathrm{h}}/\alpha_{\mathrm{c}}$ becomes
larger.
Thus, in model 9 as well as model 10 the part of the disc being
truncated during quiescence becomes larger (see Fig.\,\ref{f-cal3}).

Conversely, the accretion rate on the upper branch
which corresponds to $\Sigma_{\rm max}$ is increasing with
$\alpha_{\mathrm{h}}/\alpha_{\mathrm{c}}$.
Thus, after the outside--in heating front passes the inner disc regions,
the accretion rate there is higher for model 9 and 10 reducing the predicted
UV-delay. In the case of model 9 the UV delay is additionally shortened as
the higher value of $\alpha_{\mathrm{h}}$ leads to faster heating fronts (see
Fig.\,\ref{f-cal3}).

Considering the EUV delay we find drastic changes only in the case of
increased $\alpha_{\rm h}$ (compare model 8 and model 9). This shows that
the EUV delay is mainly governed by the viscous time scale. Due to the 
increase of $\alpha_{\rm h}$ it takes the disc less time to reach
high accretion rates.

\subsection{Variations of the mass transfer rate}

So far we presented light curves assuming the accretion disc
has adjusted to the prevailing constant mass transfer rate and is
going through the same outburst cycle all the time. Inspecting
Fig.\,\ref{f-sscvis} indicates that for SS\,Cyg this is not the
case: for example the first as well as the forth panel show
extremely irregular outburst behaviour.
In addition, no constant mass-transfer rate model can reproduce
alternating outside-in and inside-out outbursts.

Considering the obvious irregularities in the light curve we
should take into account mass transfer variations. Recently there
has been a debate about the strength of possible mass transfer
variations in dwarf novae. It has been shown that Z\,Cam light
curves can be explained assuming only small fluctuations
($\sim30\%$) of the mass transfer rate if additional heating of
the outer edge is considered \citet{buat-menardetal01-2}. On the
other hand the dwarf nova system RX\,And has occasionally
much stronger mass transfer variations
and is blend of a Z\,Cam star and VY\,Scl star
\citep{schreiberetal02-1,hameury+lasota02-1}. However, because the long term
light curve of SS\,Cygni is in general rather ``regular'' we assume
that variations of the mass transfer rate in SS\,Cyg do not exceed
the strength necessary to explain Z\,Cam standstills \footnote{In fact
SS\,Cygni is expected to show the Z\,Cam phenomenon
\citep{buat-menardetal01-2}} and use
\begin{equation}
\Mtr=\Mtr_{,0}(1+0.15\sin(\pi\,t_{100})),
\end{equation}
where $t_{100}$ is the time in units of $100$\,days. This
prescription for variations of the mass transfer rate is obviously
arbitrary but, nevertheless, the numerical experiment might
give us a hint about the nature of mass transfer variations in
SS\,Cyg.

The disc reacts on the varying mass transfer rate with alternating
long and short outbursts. This is not surprising as earlier
calculations \citep[e.g.][]{king+cannizzo98-1,schreiberetal00-1}
have shown that the disc rather quickly adjusts itself to a given
mass transfer rate. Our calculations confirm the result of
\citet{buat-menardetal01-1} that small mass transfer variations
can lead to alternating inside--out and outside--in outbursts.
Moreover, the long term light curve contains long and short
outbursts of both types. The resulting delays are given in
Table\,\ref{t-sample} (model 11).
Evidently, smooth, periodic and
small variations of the mass transfer rate do not change anything
as they do neither essentially affect the properties of the
heating fronts nor influence the viscous time scale.

\section{Comparison with earlier calculations \label{s-prev}}

As the UV delay has been discussed intensively in the literature
we should relate our results to previous calculations. After
observational evidence for the existence of the UV delay
established several papers stating that the predicted delay is too
short appeared \citep[e.g.][]{cannizzo+kenyon87-1}. The reasons
for this and later similar assertions are discussed in detail in
the excellent paper by \citet{smak98-1}.

In this paper Smak presented a model similar to the one specified
in Sect.\,\ref{s-model}.
Having calculated light
curves for a set of parameters \citet{smak98-1} concluded
that the DIM predicts correct UV delays if one uses the
correct boundary conditions and sufficiently large discs: the
delay just depends on whether an outside--in (long delay) or an
inside--out outburst (short delay) develops.

\begin{center}
\begin{table}
\caption[]{\label{t-smak} The predicted UV delay when using the same orbital
parameter as \citet{smak98-1}.
Model 1s and 2s are calculated using our model described in Sect.\ref{s-model}
whereas in model 3s and 4s we used Smak's treatment of the boundary layer
emission, and neglected (as he does) additional heating of the outer disc
and irradiation of the secondary.}
\setlength{\tabcolsep}{1.8ex}
\begin{tabular}{lllllr}
\hline\noalign{\smallskip} 
\hline\noalign{\smallskip} 
model &
\multicolumn{1}{c}{$\dot{M}_{\mathrm{tr}}$}&
\multicolumn{1}{c}{$\Delta_{\mathrm{UV,0}}$} &
\multicolumn{1}{c}{$\Delta_{\mathrm{UV,0.5}}$}&
\multicolumn{1}{c}{$R_{\mathrm{init}}$} &
\multicolumn{1}{c}{W} \\
&
\multicolumn{1}{c}{[$10^{16}{\rm gs}^{-1}$]}&
\multicolumn{1}{c}{[d]} & \multicolumn{1}{c}{[d]}&
\multicolumn{1}{c}{[$10^{10}$\,cm]}&
\multicolumn{1}{c}{[d]}\\
\noalign{\smallskip}\hline\noalign{\smallskip}
1s & 8       & 0.9 & 0.6 & 0.075 & 22  \\
2s & 10      & 0.1 & 0.5 & 1.2 & 37   \\
3s & 15      & 1.1 & 1.2 & 0.078 & 9   \\
4s & 30      & 0.4 & 1.1 & 2.44 & 10   \\
\noalign{\smallskip}\hline\noalign{\smallskip}
\end{tabular}\linebreak
\end{table}
\end{center}

\begin{figure*}
\begin{center}
\includegraphics[width=8.3cm, angle=270]{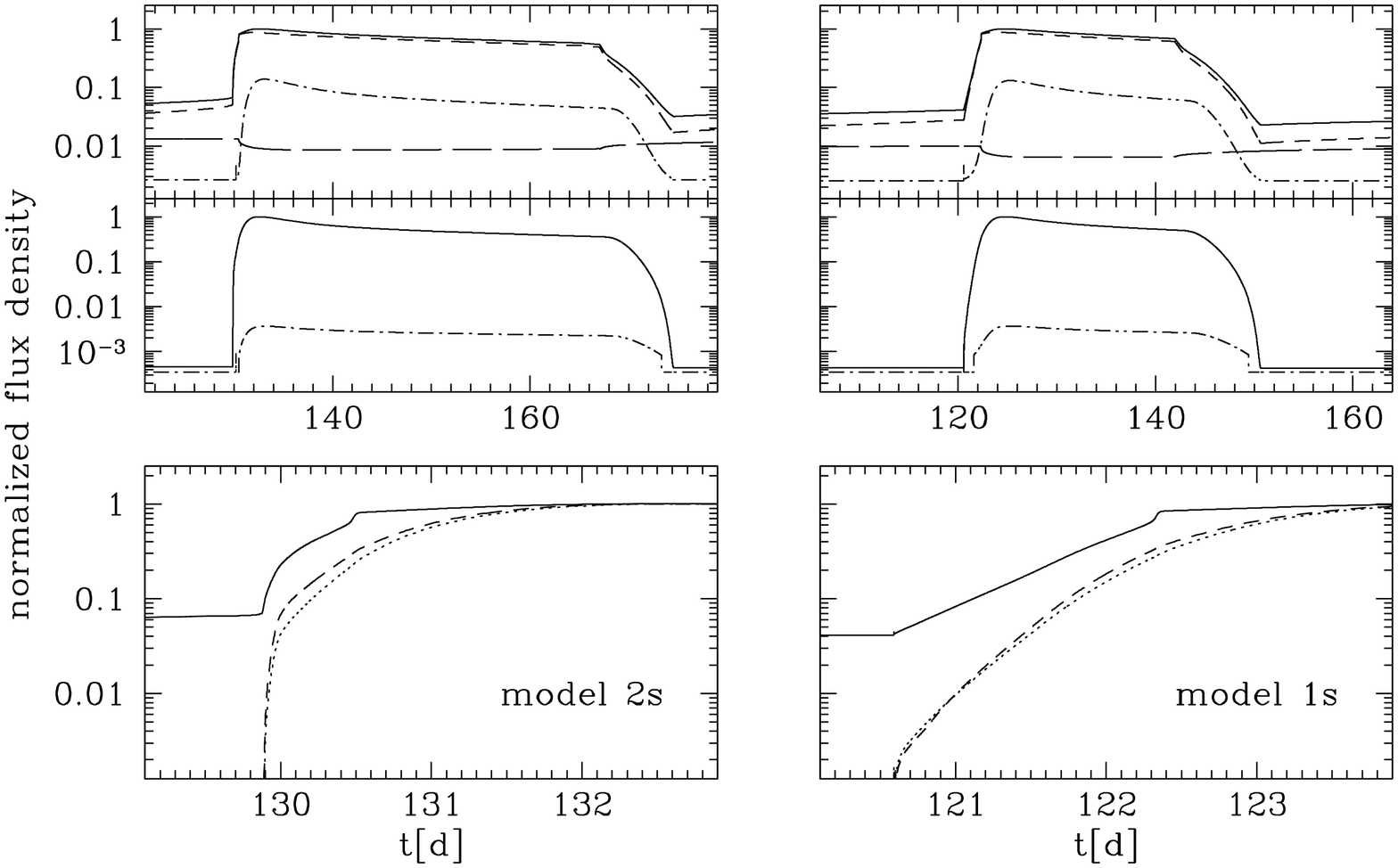}
\caption{\label{f-smak1} Comparison of inside--out (right) and outside--in
(left) outbursts calculated using the same parameter as
\citet{smak98-1}. The top panel compares the contribution to the optical flux
(solid line) from disc (short dashed), irradiated secondary
(dashed--dotted), and bright spot (long dashed). For the UV flux only emission
from the disc (second panel, solid line) is important. The boundary layer
contribution (dashed dotted) is less than one percent. The bottom panel shows
the rise of the outbursts for the optical flux (solid line) and the UV flux
assuming Kurucz (dashed) and blackbody (dotted) spectra.}
\includegraphics[width=8.3cm, angle=270]{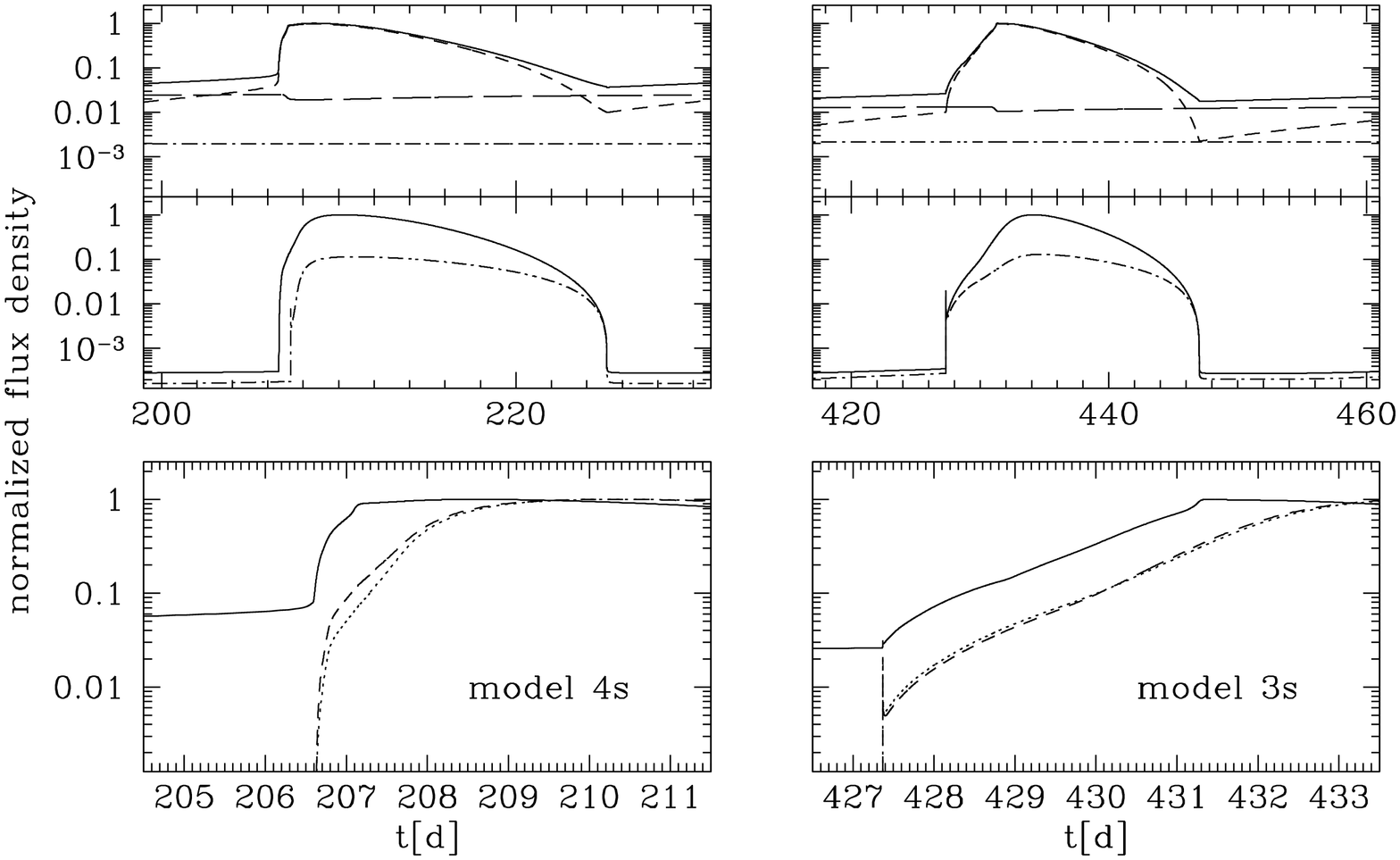}
\caption{\label{f-smak2} The same as Fig.\ref{f-smak1} but neglecting
heating due to stream impact and tidal dissipation and using
Smak's approximation for the emission from the boundary layer and the
secondary (see text).
Although the ignition radius is larger for the outside--in outburst ($R_{\rm
init}=2.4\times\,10^{10}$\,cm)
the UV delay (bottom panels) remains somewhat longer for outside--in
outbursts (see also Table\,\ref{t-smak}).}
\end{center}
\end{figure*}

However, \citet{smak98-1} did not consider heating of the
disc due to the stream impact and tidal dissipation. Without
these additional physical effects Smak was unable to obtain
outside-in (Type A - in his terminology) outbursts for a binary
with SS\,Cyg parameters: the required high mass-transfer rates
would correspond to steady accretion.
\citet{smak98-1} therefore used binary
parameter different from those listed in
Table\,\ref{t-ssc_par}, which were not directly applicable to
SS Cyg; in fact his discs are too small for this system.
In addition we note that Smak used another approximation for the emission
of the boundary layer
and the accretion disc during quiescence.

To compare our results with Smak's findings we again calculate
outburst using the model described in Sect.\,\ref{s-model} but
assuming $\Mwd=1\,\Msun$, $\Msec=0.4\,\Msun$, $\Porb=4.2$\,hr,
$<R_{\rm out}>=4.2\times\,10^{10}$\,cm and
$\Mtr=8,10\times10^{16}$\,g\,s$^{-1}$ (model 1s and 2s in
Table\,\ref{t-smak} which corresponds to model 13 and 14 of
\citet{smak98-1}). We also performed simulations without including
heating due to the stream disc impact and tidal dissipation,
assuming that the boundary layer luminosity is thermalized over
the surface of the white dwarf, i.e. $f_{\rm em}=1$ in
Eq.\,(\ref{eq:twd}), and neglecting irradiation of the secondary
(model 3s and 4s in Table\,\ref{t-smak}). Apart from the different
treatment of the lower branch of the S-curve model 3s and 4s are
identical to Smak's simulations. Due to the absence of additional
heating of the outer disc we require higher values for the mass
transfer rate to produce large outbursts, i.e.
$\Mtr=1.5\times10^{17}$g\,s$^{-1}$ and
$\Mtr=3.0\times10^{17}$g\,s$^{-1}$. The resulting delays are
listed in Table\,\ref{t-smak}.

Fig.\,\ref{f-smak1} compares the contribution of disc, boundary
layer, white dwarf, hot spot, and irradiated secondary for model s1 and s2.
Emission from the disc dominates at UV as well as
optical wavelengths. The irradiated secondary accounts for $\sim\,10\%$ of the
visual emission.  
In contrast, using Smak's prescription (model 3s and 4s; Fig.\,\ref{f-smak2})
the boundary layer contributes significantly to the UV 
(middle panel in
Fig.\,\ref{f-smak2}) and neglecting irradiation of the secondary reduces the
predicted optical emission (top panel of Fig.\,\ref{f-smak2}). In addition, 
ignoring heating of the outer
disc allows us to obtain outside--in outbursts which are triggered at
essentially larger radii, i.e. $R_{\rm init}=2.4\times10^{10}$\,cm.
Due to these changes the UV-delay becomes longer (see Table\,\ref{t-smak}).

Considering outside--in and inside--out outbursts 
our previous findings are confirmed: the
UV delay is {\em longer} for inside-out outbursts at the onset of
the optical rise and comparable close to the maximum.
We note that in Smak's calculations the instability is triggered at larger
radii. In model 13 of \citet{smak98-1} the heating front starts at
$R_{\rm init}=3.5\times\,10^{10}$\,cm whereas we do not obtain a
comparable large ignition radius ($R_{\rm init}$ in Table\,\ref{t-smak}).
This is probably due to a different treatment of the
lower branch of the S curve (Smak, private communication).

In Fig.\,\ref{f-smak1} and \ref{f-smak2} we also plotted
the UV light curve calculated by summing blackbody spectra. In agreement
with Smak's finding, the UV blackbody light curves are somewhat delayed
($\sim\,0.01-0.1$\,d) to the Kurucz ones.

\citet{hameuryetal99-1} calculated the EUV delay for SS Cyg
parameters, assuming that EUV emission
is directly proportional to the accretion rate onto the white
dwarf when the disc reaches its surface, i.e. when a
boundary layer forms. They also assume that truncation of the inner disc
but due to evaporation instead of a magnetic field
However, their inner disc radius in quiescence is comparable to the one used
here and despite of the different assumptions, the EUV delay they obtain
(1 day) is in good agreement with the value found here, showing that the
detailed mechanism causing the disc truncation is of little
importance.

Regarding the case study of \citet{cannizzo01-1} we stress the same difference
as to Smak's calculations:
even for outside--in outbursts the heating front does not start close to the
outer edge but at a distance of $\sim1-2\times\,10^{10}$\,cm from the white
dwarf.

A final note concerns the EUV: both authors
\citep{cannizzo01-1,smak98-1} assume that the EUV delay is almost
identical to the time it takes the heating front to reach the
inner edge of the disc. Our model takes recent X-ray observations
(see Sect.\,\ref{s-obs}) into account which indicate that this is
not true. Moreover, our calculations predict that
the EUV delay close to maximum is governed by the time it takes the disc to
enter into the quasi--stationary state whereas the delays at the
beginning of the optical rise also depend on properties of the
heating front and \Mcr.

\section{Discussion: model versus observations \label{s-comp}}
In this section we compare our results with the observations
outlined in Sect.\,\ref{s-obs}. It is {\em not} however
the goal of this paper to try to reproduce the observations by
using the whole parameter space.
Instead, we analyzed
systematic dependencies of the model on rather unconstrained
parameter to see whether comparison with observations may help to
guess their values.

\subsection{X-rays}

There has been only one SS Cyg outburst which was
observed simultaneously in X-rays and EUV. The observed long
delay (0.9 days) for the rise of the X-rays indicates
that this outburst was of the outside--in type as our calculations
predict no delay for inside--out outbursts. However, the X-ray
delay predicted by the calculated outside--in outbursts is shorter
than the observed one by a factor of $\sim$\,2--3.

The obtained durations of the ``X-ray on state'' during the
optical rise ($\sim\,0.5$\,days) as well as at the end of the
outburst ($\sim\,2$\,days) are in perfect agreement with the
observed ones \citep[see][]{wheatley00-1,wheatleyetal03-1}.

Comparing the observed X-ray flux during quiescence with the
calculated mass accretion rate strongly suggests that the inner
disc is truncated. Truncation leads to {\em higher} accretion
rates during quiescence as accretion increases with radius while
the disc accumulates mass.

Concerning the X-ray delay and the duration of the ``X-ray on'' state
one should note that our results are obviously sensitive to $\Mcr$.

\subsection{The UV delay}

Generally, the delay between the optical and the UV rise is
comparable to the observed one (which has always been found to
be of order of half a day, see above) for model 1--8, and 11 (see
Table\,\ref{t-obs1} and \ref{t-sample}). Although the delay is
somewhat {\em longer} for inside--out outbursts, the agreement
between observations and calculations is independent on whether
the heating front starts close to the white dwarf or not. Thus,
the UV delay is {\em not} a good indicator of the
outburst--type.

The UV delay sensitively depends on $\alpha$ (see
Sect.\,\ref{s-alpha}). We obtain good agreement with $\alpha_{\rm h}=0.1$,
and $\alpha_{\rm c}=0.02$.

\subsection{The EUV delay}

\subsubsection{Normal outbursts}

At first we consider the so called normal outbursts (see
Fig.\,\ref{f-sscvis} and Table\,\ref{t-obs1}). These outbursts are
characterized by a fast optical rise and are therefore
generally thought to be of the outside--in type.
We find the calculated EUV delay at half the maximum optical flux
($\Delta_{\rm EUV,0.5}$) being nearly independent on the radius at
which the heating front was triggered (Table\,\ref{t-sample}) but
in excellent agreement with the observed delay. Our model predicts
a somewhat lower value at the onset of the optical rise
($\Delta_{\rm EUV,0}$) than it is observed except for inside--out
outbursts without truncation. We note that $\Delta_{\rm EUV,0}$
sensitively depends on the assumed value of $\Mcr$ and the optical
flux in quiescence (the effective temperature of the secondary
might be higher than 4000\,K as suggested e.g. by
\citet{webbetal02-1}). Of course, it would be possible to
perfectly match the observed $\Delta_{\rm EUV,0}$ also in the case
of outside--in outbursts with truncation. To sum up, the model
can reproduce the EUV delay measured for normal outbursts.

\subsubsection{Anomalous outbursts}

The standard $\alpha$-model, even modified by truncation,
various types of heating etc., cannot reproduce the observed light curve
of anomalous outbursts. 
The model does never predict a rise time comparable to
that of the outburst observed in 1999 ($\sim17$\,d; Fig.\,\ref{f-sscvis}
bottom panel). The observed rise time of the anomalous outburst in 
1993 ($\sim\,5$\,d; Fig.\,\ref{f-sscvis} second panel from top) can only be 
reproduced if we assume that the disc extends down to the surface of the 
white dwarf (i.e. no truncation) and if we additionally completely
neglect both heating of the outer disc due to the stream impact and tidal
dissipation. In this case, however, the calculated
light curve does not have the observed flat top which is as bad as the
prediction of a too short rise time.
In addition, without truncation the model predicts several small outbursts 
between the larger ones which is not observed (Fig.\,\ref{f-cal6}), 
and also the predicted X-ray flux is far too small (see Sect.\,4.1 and 4.2).

In view of this incapacity of the model, trying to find the corresponding
delays appears totally pointless. However, attempting to find models
with time lags close to those observed could be a useful
diagnostic tool for future developments.

As mentioned, we have to distinguish 
between the 1993 (second panel Fig.\,\ref{f-sscvis}) and the 1999 (bottom 
panel) event as their properties are very different. Considering the 1999 
eruption we note that the long delay between the optical and the EUV measured
at the onset of the optical rise of the anomalous outbursts
(Table.\,\ref{t-obs1}) is not reproduced even by rather long
inside--out outbursts without truncation (i.e. model 3). Assuming
the inner disc being truncated, the situation becomes even worse
as $\Delta_{\rm EUV,0}$ gets shorter. In addition, our simulations
disagree with the observed drastic decrease of the EUV delay with
increasing optical flux
(Table\,\ref{t-obs1},\citet{maucheetal01-1}). It is especially the
observed slow optical rise and not the shape of the EUV light
curve which differs from the predictions. Inspecting
Fig.\,\ref{f-sscvis} we find the outburst being a part of strong
deviations from regular outburst behaviour: there was no
quiescence at all between this outburst and the previous one. It
is then not at all surprising that the model cannot reproduce the
EUV delay of this particular outburst. Such a disruption of the
light curve can only be due to a change in one of the parameters
assumed here to be constant. The mass transfer rate could for
example have changed drastically, or the Shakura-Sunyaev
parameterization of the viscosity could have been affected by e.g.
a magnetic flare in the disc; there are many possibilities, and it
is far beyond the scope of this paper to investigate them in
detail.

For the 1993 outburst we note that our calculations hardly agree
with the long delay at the onset of the optical rise. Especially
if the inner disc is truncated (which is strongly suggested by the
quiescence X-ray flux) the obtained value for $\Delta_{\rm EUV,0}$
is smaller than observed even for inside--out outbursts whereas
for the predicted time lags closer to the maximum the opposite
holds. Of course, $\Delta_{\rm EUV,0}$ still depends on the
approximation of the boundary layer (e.g. \Mcr) and the optical
emission during quiescence (secondary). We certainly would be able to
construct a delay of three days but in any case the model
predicts a significant delay closer to the maximum. This
prediction of the model is firm as the EUV emission reaches
maximum when the disc has adjusted to the quasi-stationary state
whereas the optical flux is already close to maximum when the
heating front has gone through the outer disc region. However,
the obtained disagreement could be related to the three
low-amplitude anomalous outbursts preceding by $\sim\,20$\, days
the anomalous 1993 outburst.

\subsection{The EUV spectrum}
\citet{maucheetal95-1} present EUVE observations of SS\,Cyg during the
outburst in 1993 between August 17.1 and 23.6. They find a remarkable
correlation between the hard (72--90\,\AA) and soft (90--130\,\AA)
count rate. Even during the rise of the outburst the ratio of the two
remains constant.
On the other hand, more recent observations of another outburst
show that the hardness ratio may change by a factor of $\sim\,3$
\citet{wheatleyetal03-1}. Although our assumptions for the
boundary layer are quite rough, we calculate the hardness ratio
predicted by the model to get a hint of  the
uncertainty linked to the black body approximation.

\begin{figure}
\includegraphics[width=8.5cm, angle=0]{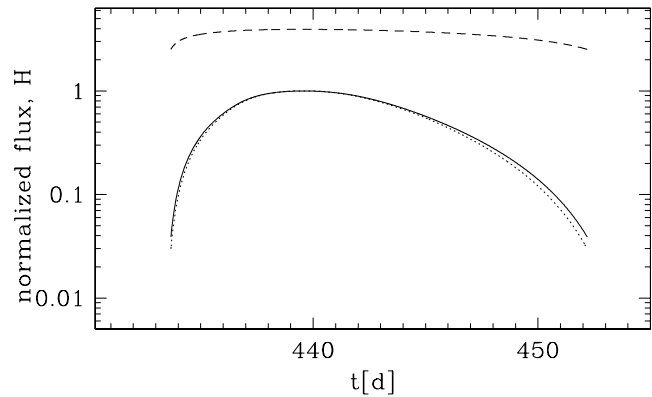}
\caption[]{\label{f-cal7} Normalized monochromatic flux at 100\,\AA\ (solid
line), integrated EUV flux (70--130\,\AA; dotted line), and the EUV hardness
ratio (dashed line). }
\end{figure}

Assuming $N_{\rm H}=4.4\,\times\,10^{19}$\,cm$^{-2}$, making use of the
cross-sections as a function of wavelength according to
\citet{morrison+mccammon83-1} by interpolating their Table\,2, we
simulated the evolution of the EUV spectrum during an outburst.
Fig.\,\ref{f-cal7} shows the normalized integrated flux
(70--130\AA), monochromatic $100$\AA\ flux, and the hardness ratio
defined as $H$ = flux(70--90\AA)/flux(90--130\AA)\, for model 1. We
find $H$ varying between 2 and 4 which appears to be in reasonable 
agreement with the count rate hardness ratio given by \citet{wheatley00-1} 
and \citet{maucheetal95-1}.

\subsection{Bandpass fluxes}
So far we have presented monochromatic model light curves  
and compared the obtained delays with the observed time lags.
Here we verify that our monochromatic approach is reasonable. 

Fig.\,\ref{f-cal7} confirms that the 
EUV light curve 
is well represented by the 100\AA\, emission. 
The suitability of the monochromatic light curves 
is further established in Fig.\,\ref{f-cal8} where we present 
both the predicted normalized flux densities at 5500\AA\, and 100\AA\, 
(top panel) and the corresponding integrated optical and EUV fluxes 
(bottom panel). 
Apparently, the light curves and especially the
predicted delays are very similar. The only difference worth
mentioning concerns the optical emission: 
compared to the monochromatic light curves the relative contribution of 
the hot spot emission increases, which leads to relatively 
brighter quiescence. 
Concerning the delays we conclude that the monochromatic light 
curves can be considered as representative for the emission 
in the corresponding bandpasses. 

\subsection{Absolute values}

Discussing the absolute emission in addition to the  
normalized monochromatic light curves requires to assume 
a distance for SS\,Cygni. 
Recently \citet{harrisonetal99-1} 
measured a parallax using the HST Fine Guiding Sensor (FGS) 
and derived a distance of
$d=166\pm12$\,pc which is essentially larger than previously thought.
We want to stress here that, according to the model, the distance of SS\,Cygni
cannot be $166$\,pc. In all the calculations performed for
this paper the maximum accretion rate hardly reaches
$\Macc=10^{18}$g\,s$^{-1}$ which is far too low to explain the observed 
visual brightness of the system if its distance indeed was $166$\,pc. 
As shown by \citet{schreiber+gaensicke02-1}, 
at such a large distance SS\,Cyg would not be dwarf nova: the accretion rate
required to reproduce the observed visual brightness would certainly 
lead to stationary accretion \footnote{It is worth noting that
problems with the small HST/FGS--parallax have been mentioned by 
\citet{northetal02-1} too. The space velocity they derive
assuming $d=166$\,pc is essentially larger than theoretically expected.}.  

\begin{figure}
\includegraphics[width=9.0cm, angle=0]{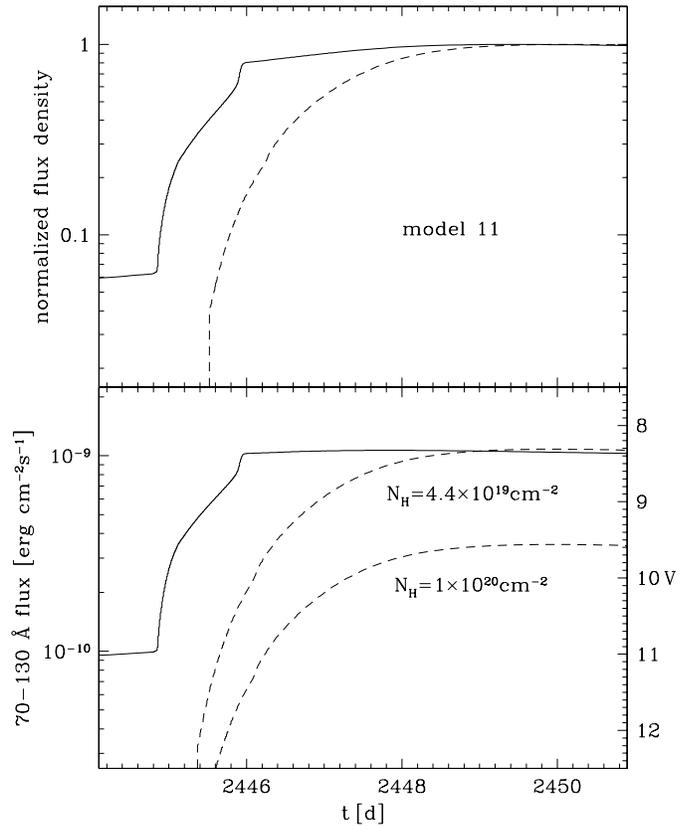}
\caption[]{\label{f-cal8} 
Light curves of the long outside--in outburst 
we obtain assuming small variations of the mass transfer rate (model 11).
Top panel: normalized flux density at 5500\,\AA\, (solid line) and 
100\,\AA\, (dashed line). Bottom panel: visual magnitude and integrated 
70--130\,\AA\, flux assuming $d=100$\,pc. }
\end{figure}

Instead of the large HST/FGS distance we therefore 
assume a more traditional and, according to the model, more realistic 
value for the distance of SS\,Cygni, i.e. $d\sim\,100$\,pc
\citep[e.g.][]{warner87-1,bailey81-1,kiplinger79-1}. 
The order of the maximum monochromatic emission of the light curves 
presented in Figs.\,\ref{f-cal1}--\ref{f-cal3} is the following: 
the lowest flux density is
expected at the optical wavelength   
($\sim\,1.5\times\,10^{-14}$\,erg\,cm$^{-2}$s$^{-1}$\,\AA$^{-1}$)   
followed by the UV 
($\sim\,3.3\times\,10^{-13}$\,erg\,cm$^{-2}$s$^{-1}$\,\AA$^{-1}$)
and the EUV
($\sim\,2-12\times\,10^{-11}$\,erg\,cm$^{-2}$s$^{-1}$\,\AA$^{-1}$). 
These values are calculated assuming an inclination of $i=37^°$. 
The EUV emission depends strongly on the assumed hydrogen column 
density. The range of values given above is
determined assuming  $N_{\rm H}$ being between $10\times10^{19}$\,cm$^{-2}$ 
and $4.4\times10^{19}$\,cm$^{-2}$.
In Fig.\,\ref{f-cal8} we also give absolute band integrated optical and 
EUV ($70-130$\AA\,) fluxes.  

Comparison with Fig.\,\ref{f-sscvis} shows that 
the obtained maximum visual magnitude agrees well with the observations. 
Furthermore, our calculations predict that the  
maximum UV flux density exceeds the optical one by a factor $\sim\,30$ 
which is also in very good agreement with 
the observations \citep{cannizzoetal86-1}. 
The calculated absolute EUV emission is certainly the most 
uncertain one as it does not only sensitively depend on 
$N_{\rm H}$ but also on uncertainties related to the temperature and 
fractional emitting region of 
the boundary layer. Indeed, taking these difficulties into account, 
it has been shown by \citet[][]{wheatleyetal03-1} that the range of 
blackbody luminosities for which the observed EUV spectrum can be fitted 
is uncertain by more than one magnitude. 
Hence, trying to constrain the model using absolute EUV fluxes seems not 
very promising. Nevertheless, for completeness we note rough agreement 
between our calculations and the observations 
\citep[see e.g.][]{maucheetal95-1,mauche02-1} 
for large hydrogen column densities, i.e. $N_{\rm
H}\,\gappr\,10^{20}$\,cm$^{-2}$.

\subsection{The long term light curve}

Although the main goal of this paper is to analyze the time lags
in SS\,Cyg, the predicted long term light curve is an
essential feature which must agree with observations for the
parameters that best explain the delays. The outburst cycles
depend on the details of modeling. The parameter we chose
lead to three different cycles (Fig.\,\ref{f-cal6}). With the disc
extending down to the white dwarf, we get several small
inside--out outburst between the major eruptions that are not
observed. Truncation of the inner disc suppresses these small
outbursts and the light curve consists only of large outbursts. If
in addition small fluctuations of the mass transfer rate are
included, it is possible to obtain alternating short and long
outbursts of both types. Obviously, the latter hypothesis is
favoured. Getting irregularities comparable to those seen in the observed
light curve (see e.g. second and last panel from the top in
Fig.\,\ref{f-sscvis}) requires a major revision of the model.

\section{Conclusion}
We present a physically realistic model for dwarf novae consisting of the
disc instability model \citep{hameuryetal98-1} and simple but reasonable
assumptions for the emission of the boundary layer.
We calculated dwarf nova light curves using parameters appropriate for
SS\,Cygni to investigate time lags observed between the rise at different
wavelengths.
The results from this study are:

\begin{enumerate}

\item The UV delay strongly depends on where one measures it.
Close to maximum it is nearly independent of where the heating
front started or whether the inner disc is truncated. The UV delay
at the beginning of the optical rise is significantly {\em longer}
when the heating front is triggered close to the white dwarf.

The UV delay is not at all a good indicator for the outburst type.
and it is sensitive to the value of $\alpha$.
We find reasonable agreement with observations
for $\alpha_{\rm c}=0.02$ and $\alpha_{\rm h}=0.1$.

\item The agreement between observed and predicted EUV delay is
satisfying for normal outbursts whereas the model fails to
reproduce anomalous outbursts.

\item The increased X-ray flux observed between the optical and
the EUV rise is a natural outcome of outside--in outbursts. At the
end of every outburst the model predicts a rise of the X-rays
comparable to the observed one. X-ray observations during
quiescence strongly support the idea of truncation of the inner
disc.

\end{enumerate}

\begin{acknowledgements}
We are grateful to Chris Mauche for very helpful comments and advice.
This research has made use of the AFOEV database, operated at CDS,
France. MRS acknowledges funding by an individual Marie--Curie fellowship. 
This work has benefited from developments made by
V.Buat-M\'enard during his PhD thesis. JPL thanks 
Chris Mauche for enlightening discussions about various delays.
\end{acknowledgements}


\end{document}